\documentstyle[10pt,epsf,epsfig,hangcaption,xspace,amssymb,amsfonts,amsmath,amsthm,cite,
dp_delphititle,lineno]{dp_delphi}
\makeindex
\def\DpPaperGroup{EP}
\def\DpPaperRef{2003-048}
\def\DpDate{14 July 2003}
\def\DpAuthors{DELPHI Collaboration}
\def\DpTitle{{Search for SUSY in the AMSB scenario with the DELPHI detector}}
\def\DpSubmit{(Accepted by Eur. Phys. J. C)}
\def\DpComment{  }
\def\DpEMail{ }

\def\leqsim{\mathbin{\;\raise1pt\hbox{$<$}\kern-8pt\lower3pt\hbox{$\sim$}\;}}
\def\geqsim{\mathbin{\;\raise1pt\hbox{$>$}\kern-8pt\lower3pt\hbox{$\sim$}\;}}

\def\MXN#1{\mbox{$ M_{\tilde{\chi}^0_#1}                                $}}
\def\MXC#1{\mbox{$ M_{\tilde{\chi}^{\pm}_#1}                            $}}
\def\XP#1{\mbox{$ \tilde{\chi}^+_#1                                     $}}
\def\XM#1{\mbox{$ \tilde{\chi}^-_#1                                     $}}
\def\XPM#1{\mbox{$ \tilde{\chi}^{\pm}_#1                                $}}
\def\XN#1{\mbox{$ \tilde{\chi}^0_#1                                     $}}

\def\p#1{\mbox{$ \mbox{\bf p}_1                                         $}}
\def\mchi{\mbox{$ M_{\tilde{\chi}}                                      $}}

\newcommand{\tanb}    {\mbox{$ \tan \beta                                  $}}
\newcommand{\mzero}   {\mbox{$ m_0                                         $}}
\newcommand{\mhf}     {\mbox{$ m_{3/2}                                     $}}
\newcommand{\sgnmu}   {\mbox{$ {\mathrm sign}(\mu)                         $}}
\newcommand{\smu}     {\mbox{$ \tilde{\mu}                                 $}}

\newcommand{\sel}     {\mbox{$ \tilde{\mathrm e}                           $}}

\newcommand{\snu}     {\mbox{$ \tilde\nu                                   $}}
\newcommand{\msnu}    {\mbox{$ M_{\tilde\nu}                               $}}

\newcommand{\sle}     {\mbox{$ \tilde{l}                                   $}}
\newcommand{\msle}    {\mbox{$ M_{\sle}                                    $}}
\newcommand{\stau}    {\mbox{$ \tilde{\tau}                                $}}
\newcommand{\stone}   {\mbox{$ \tilde{\tau}_1                              $}}

\newcommand{\mstone}  {\mbox{$ M_{\tilde{\tau}_1}                          $}}

\newcommand{\hn}      {\mbox{$ {\, \mathrm h}^0                               $}}
\newcommand{\Zn}      {\mbox{$ {\, \mathrm Z}                                 $}}

\newcommand{\ee}      {\mbox{$ {\, \mathrm e}^+ {\mathrm e}^-                 $}}

\newcommand{\eeto}    {\mbox{$ {\, \mathrm e}^+ {\mathrm e}^- \to             $}}

\newcommand{\MeVcc}   {\mbox{$ {\mathrm{MeV}}/c^2                          $}}

\newcommand{\GeVcc}   {\mbox{$ {\mathrm{GeV}}/c^2                          $}}

\newcommand{\TeVcc}   {\mbox{$ {\mathrm{TeV}}/c^2                          $}}
\newcommand{\pbi}     {\mbox{$ {\mathrm{pb}}^{-1}                          $}}
\newcommand{\MZ}      {\mbox{$ M_{{\mathrm Z}}                             $}}

\newcommand{\MA}      {\mbox{$ M_{\mathrm A}                               $}}

\newcommand{\MSH}     {\mbox{$ M_{{\mathrm h}^0}                           $}}

\newcommand{\etal}  {\mbox{\it et al.}}
\def\NPB#1#2#3{{\rm Nucl.~Phys.} {\bf{B#1}} (#2) #3}
\def\PLB#1#2#3{{\rm Phys.~Lett.} {\bf{B#1}} (#2) #3}
\def\PRD#1#2#3{{\rm Phys.~Rev.} {\bf{D#1}} (#2) #3}
\def\PRL#1#2#3{{\rm Phys.~Rev.~Lett.} {\bf{#1}} (#2) #3}

\def\NIMA#1#2#3{{\rm Nucl.~Instr.~and~Meth.} {\bf{A#1}} (#2) #3} 
\def\CPC#1#2#3{{\rm Comp.~Phys.~Comm.} {\bf#1} (#2) #3}
\def\EPJC#1#2#3{{\rm Eur.~Phys.~J.} {\bf{C#1}} (#2) #3}

\def    \DM          {\mbox{$\Delta M$}}
\def    \DSNU        {\mbox{$\Delta M_{\tilde\nu}$}}
\def    \DSLE        {\mbox{$\Delta M_{\tilde\ell}$}}
\def    \rs          {\mbox{$\sqrt{s}$}}
\def    \lik         {\mbox{$\mathcal L_R$}}
\def    \likcut      {\mbox{$\mathcal L_{R_{\mathrm CUT}}$}}

\begin{document}
\makeatletter
\newcount\@tempcntc
\def\@citex[#1]#2{\if@filesw\immediate\write\@auxout{\string\citation{#2}}\fi
  \@tempcnta\z@\@tempcntb\m@ne\def\@citea{}\@cite{\@for\@citeb:=#2\do
    {\@ifundefined
       {b@\@citeb}{\@citeo\@tempcntb\m@ne\@citea\def\@citea{,}{\bf ?}\@warning
       {Citation `\@citeb' on page \thepage \space undefined}}%
    {\setbox\z@\hbox{\global\@tempcntc0\csname b@\@citeb\endcsname\relax}%
     \ifnum\@tempcntc=\z@ \@citeo\@tempcntb\m@ne
       \@citea\def\@citea{,}\hbox{\csname b@\@citeb\endcsname}%
     \else
      \advance\@tempcntb\@ne
      \ifnum\@tempcntb=\@tempcntc
      \else\advance\@tempcntb\m@ne\@citeo
      \@tempcnta\@tempcntc\@tempcntb\@tempcntc\fi\fi}}\@citeo}{#1}}
\def\@citeo{\ifnum\@tempcnta>\@tempcntb\else\@citea\def\@citea{,}%
  \ifnum\@tempcnta=\@tempcntb\the\@tempcnta\else
   {\advance\@tempcnta\@ne\ifnum\@tempcnta=\@tempcntb \else \def\@citea{--}\fi
    \advance\@tempcnta\m@ne\the\@tempcnta\@citea\the\@tempcntb}\fi\fi}
 
\makeatother
\begin{titlepage}
\pagenumbering{roman}
\CERNpreprint{\DpPaperGroup}{\DpPaperRef} 
\date{{\small\DpDate}} 
\title{\DpTitle} 
\address{\DpAuthors} 
\begin{shortabs} 
\noindent
%
The DELPHI experiment at the LEP \ee\ collider collected almost 700~\pbi\
at centre-of-mass energies above the $Z^0$ mass pole and up to 208 GeV.
Those data were used to search for SUSY in the Anomaly Mediated SUSY
Breaking (AMSB) scenario with a flavour independent common sfermion mass
parameter.
The searches covered several possible signatures experimentally accessible
at LEP, with either the neutralino, the sneutrino or the stau being the 
Lightest Supersymmetric Particle (LSP). They included: the search for 
nearly mass-degenerate chargino and neutralino, which is a typical feature 
of AMSB; the search for Standard-Model-like or invisibly decaying Higgs
boson; the search for stable staus; the search for cascade decays of SUSY
particles resulting in the LSP and a low multiplicity final state containing
neutrinos. No evidence of a signal was found, and thus constraints were set
in the space of the parameters of the model.
\end{shortabs}
\vfill
\begin{center}
\DpSubmit \ \\ 
\DpComment \ \\
\DpEMail \ \\
\end{center}
\vfill
\clearpage
\headsep 10.0pt
\addtolength{\textheight}{10mm}
\addtolength{\footskip}{-5mm}
\begingroup
%
\newcommand{\DpName}[2]{\hbox{#1$^{\ref{#2}}$},\hfill}
\newcommand{\DpNameTwo}[3]{\hbox{#1$^{\ref{#2},\ref{#3}}$},\hfill}
\newcommand{\DpNameThree}[4]{\hbox{#1$^{\ref{#2},\ref{#3},\ref{#4}}$},\hfill}
\newskip\Bigfill \Bigfill = 0pt plus 1000fill
\newcommand{\DpNameLast}[2]{\hbox{#1$^{\ref{#2}}$}\hspace{\Bigfill}}
%
\footnotesize
\noindent
\DpName{J.Abdallah}{LPNHE}
\DpName{P.Abreu}{LIP}
\DpName{W.Adam}{VIENNA}
\DpName{P.Adzic}{DEMOKRITOS}
\DpName{T.Albrecht}{KARLSRUHE}
\DpName{T.Alderweireld}{AIM}
\DpName{R.Alemany-Fernandez}{CERN}
\DpName{T.Allmendinger}{KARLSRUHE}
\DpName{P.P.Allport}{LIVERPOOL}
\DpName{U.Amaldi}{MILANO2}
% ...
\normalsize
\endgroup
\titlefoot{Department of Physics and Astronomy, Iowa State
     University, Ames IA 50011-3160, USA
    \label{AMES}}
\titlefoot{Physics Department, Universiteit Antwerpen,
     Universiteitsplein 1, B-2610 Antwerpen, Belgium \\
     \indent~~and IIHE, ULB-VUB,
     Pleinlaan 2, B-1050 Brussels, Belgium \\
     \indent~~and Facult\'e des Sciences,
     Univ. de l'Etat Mons, Av. Maistriau 19, B-7000 Mons, Belgium
    \label{AIM}}
\titlefoot{Physics Laboratory, University of Athens, Solonos Str.
     104, GR-10680 Athens, Greece
    \label{ATHENS}}
\titlefoot{Department of Physics, University of Bergen,
     All\'egaten 55, NO-5007 Bergen, Norway
    \label{BERGEN}}
\titlefoot{Dipartimento di Fisica, Universit\`a di Bologna and INFN,
     Via Irnerio 46, IT-40126 Bologna, Italy
    \label{BOLOGNA}}
\titlefoot{Centro Brasileiro de Pesquisas F\'{\i}sicas, rua Xavier Sigaud 150,
     BR-22290 Rio de Janeiro, Brazil \\
     \indent~~and Depto. de F\'{\i}sica, Pont. Univ. Cat\'olica,
     C.P. 38071 BR-22453 Rio de Janeiro, Brazil \\
     \indent~~and Inst. de F\'{\i}sica, Univ. Estadual do Rio de Janeiro,
     rua S\~{a}o Francisco Xavier 524, Rio de Janeiro, Brazil
    \label{BRASIL}}
\titlefoot{Coll\`ege de France, Lab. de Physique Corpusculaire, IN2P3-CNRS,
     FR-75231 Paris Cedex 05, France
    \label{CDF}}
\titlefoot{CERN, CH-1211 Geneva 23, Switzerland
    \label{CERN}}
\titlefoot{Institut de Recherches Subatomiques, IN2P3 - CNRS/ULP - BP20,
     FR-67037 Strasbourg Cedex, France
    \label{CRN}}
\titlefoot{Now at DESY-Zeuthen, Platanenallee 6, D-15735 Zeuthen, Germany
    \label{DESY}}
\titlefoot{Institute of Nuclear Physics, N.C.S.R. Demokritos,
     P.O. Box 60228, GR-15310 Athens, Greece
    \label{DEMOKRITOS}}
\titlefoot{FZU, Inst. of Phys. of the C.A.S. High Energy Physics Division,
     Na Slovance 2, CZ-180 40, Praha 8, Czech Republic
    \label{FZU}}
\titlefoot{Dipartimento di Fisica, Universit\`a di Genova and INFN,
     Via Dodecaneso 33, IT-16146 Genova, Italy
    \label{GENOVA}}
\titlefoot{Institut des Sciences Nucl\'eaires, IN2P3-CNRS, Universit\'e
     de Grenoble 1, FR-38026 Grenoble Cedex, France
    \label{GRENOBLE}}
\titlefoot{Helsinki Institute of Physics, P.O. Box 64,
     FIN-00014 University of Helsinki, Finland
    \label{HELSINKI}}
\titlefoot{Joint Institute for Nuclear Research, Dubna, Head Post
     Office, P.O. Box 79, RU-101 000 Moscow, Russian Federation
    \label{JINR}}
\titlefoot{Institut f\"ur Experimentelle Kernphysik,
     Universit\"at Karlsruhe, Postfach 6980, DE-76128 Karlsruhe,
     Germany
    \label{KARLSRUHE}}
\titlefoot{Institute of Nuclear Physics,Ul. Kawiory 26a,
     PL-30055 Krakow, Poland
    \label{KRAKOW1}}
\titlefoot{Faculty of Physics and Nuclear Techniques, University of Mining
     and Metallurgy, PL-30055 Krakow, Poland
    \label{KRAKOW2}}
\titlefoot{Universit\'e de Paris-Sud, Lab. de l'Acc\'el\'erateur
     Lin\'eaire, IN2P3-CNRS, B\^{a}t. 200, FR-91405 Orsay Cedex, France
    \label{LAL}}
\titlefoot{School of Physics and Chemistry, University of Lancaster,
     Lancaster LA1 4YB, UK
    \label{LANCASTER}}
\titlefoot{LIP, IST, FCUL - Av. Elias Garcia, 14-$1^{o}$,
     PT-1000 Lisboa Codex, Portugal
    \label{LIP}}
\titlefoot{Department of Physics, University of Liverpool, P.O.
     Box 147, Liverpool L69 3BX, UK
    \label{LIVERPOOL}}
\titlefoot{Dept. of Physics and Astronomy, Kelvin Building,
     University of Glasgow, Glasgow G12 8QQ
    \label{GLASGOW}}
\titlefoot{LPNHE, IN2P3-CNRS, Univ.~Paris VI et VII, Tour 33 (RdC),
     4 place Jussieu, FR-75252 Paris Cedex 05, France
    \label{LPNHE}}
\titlefoot{Department of Physics, University of Lund,
     S\"olvegatan 14, SE-223 63 Lund, Sweden
    \label{LUND}}
\titlefoot{Universit\'e Claude Bernard de Lyon, IPNL, IN2P3-CNRS,
     FR-69622 Villeurbanne Cedex, France
    \label{LYON}}
\titlefoot{Dipartimento di Fisica, Universit\`a di Milano and INFN-MILANO,
     Via Celoria 16, IT-20133 Milan, Italy
    \label{MILANO}}
\titlefoot{Dipartimento di Fisica, Univ. di Milano-Bicocca and
     INFN-MILANO, Piazza della Scienza 2, IT-20126 Milan, Italy
    \label{MILANO2}}
\titlefoot{IPNP of MFF, Charles Univ., Areal MFF,
     V Holesovickach 2, CZ-180 00, Praha 8, Czech Republic
    \label{NC}}
\titlefoot{NIKHEF, Postbus 41882, NL-1009 DB
     Amsterdam, The Netherlands
    \label{NIKHEF}}
\titlefoot{National Technical University, Physics Department,
     Zografou Campus, GR-15773 Athens, Greece
    \label{NTU-ATHENS}}
\titlefoot{Physics Department, University of Oslo, Blindern,
     NO-0316 Oslo, Norway
    \label{OSLO}}
\titlefoot{Dpto. Fisica, Univ. Oviedo, Avda. Calvo Sotelo
     s/n, ES-33007 Oviedo, Spain
    \label{OVIEDO}}
\titlefoot{Department of Physics, University of Oxford,
     Keble Road, Oxford OX1 3RH, UK
    \label{OXFORD}}
\titlefoot{Dipartimento di Fisica, Universit\`a di Padova and
     INFN, Via Marzolo 8, IT-35131 Padua, Italy
    \label{PADOVA}}
\titlefoot{Rutherford Appleton Laboratory, Chilton, Didcot
     OX11 OQX, UK
    \label{RAL}}
\titlefoot{Dipartimento di Fisica, Universit\`a di Roma II and
     INFN, Tor Vergata, IT-00173 Rome, Italy
    \label{ROMA2}}
\titlefoot{Dipartimento di Fisica, Universit\`a di Roma III and
     INFN, Via della Vasca Navale 84, IT-00146 Rome, Italy
    \label{ROMA3}}
\titlefoot{DAPNIA/Service de Physique des Particules,
     CEA-Saclay, FR-91191 Gif-sur-Yvette Cedex, France
    \label{SACLAY}}
\titlefoot{Instituto de Fisica de Cantabria (CSIC-UC), Avda.
     los Castros s/n, ES-39006 Santander, Spain
    \label{SANTANDER}}
\titlefoot{Inst. for High Energy Physics, Serpukov
     P.O. Box 35, Protvino, (Moscow Region), Russian Federation
    \label{SERPUKHOV}}
\titlefoot{J. Stefan Institute, Jamova 39, SI-1000 Ljubljana, Slovenia
     and Laboratory for Astroparticle Physics,\\
     \indent~~Nova Gorica Polytechnic, Kostanjeviska 16a, SI-5000 Nova Gorica, Slovenia, \\
     \indent~~and Department of Physics, University of Ljubljana,
     SI-1000 Ljubljana, Slovenia
    \label{SLOVENIJA}}
\titlefoot{Fysikum, Stockholm University,
     Box 6730, SE-113 85 Stockholm, Sweden
    \label{STOCKHOLM}}
\titlefoot{Dipartimento di Fisica Sperimentale, Universit\`a di
     Torino and INFN, Via P. Giuria 1, IT-10125 Turin, Italy
    \label{TORINO}}
\titlefoot{INFN,Sezione di Torino, and Dipartimento di Fisica Teorica,
     Universit\`a di Torino, Via P. Giuria 1,\\
     \indent~~IT-10125 Turin, Italy
    \label{TORINOTH}}
\titlefoot{Dipartimento di Fisica, Universit\`a di Trieste and
     INFN, Via A. Valerio 2, IT-34127 Trieste, Italy \\
     \indent~~and Istituto di Fisica, Universit\`a di Udine,
     IT-33100 Udine, Italy
    \label{TU}}
\titlefoot{Univ. Federal do Rio de Janeiro, C.P. 68528
     Cidade Univ., Ilha do Fund\~ao
     BR-21945-970 Rio de Janeiro, Brazil
    \label{UFRJ}}
\titlefoot{Department of Radiation Sciences, University of
     Uppsala, P.O. Box 535, SE-751 21 Uppsala, Sweden
    \label{UPPSALA}}
\titlefoot{IFIC, Valencia-CSIC, and D.F.A.M.N., U. de Valencia,
     Avda. Dr. Moliner 50, ES-46100 Burjassot (Valencia), Spain
    \label{VALENCIA}}
\titlefoot{Institut f\"ur Hochenergiephysik, \"Osterr. Akad.
     d. Wissensch., Nikolsdorfergasse 18, AT-1050 Vienna, Austria
    \label{VIENNA}}
\titlefoot{Inst. Nuclear Studies and University of Warsaw, Ul.
     Hoza 69, PL-00681 Warsaw, Poland
    \label{WARSZAWA}}
\titlefoot{Fachbereich Physik, University of Wuppertal, Postfach
     100 127, DE-42097 Wuppertal, Germany
    \label{WUPPERTAL}}
\addtolength{\textheight}{-10mm}
\addtolength{\footskip}{5mm}
\clearpage
\headsep 30.0pt
\end{titlepage}
%
\pagenumbering{arabic} 
\setcounter{footnote}{0} %
\large
%

\section{Introduction}

There are several theoretical motivations to believe that nature could be supersymmetric. 
However, after many years of searching in collider experiments, no evidence was found for
the existence of supersymmetric particles. The negative results of the searches constrains
the spectrum of the SUSY particles and of the parameters of the model. The mechanism of SUSY 
breaking itself is unclear. In the gravity mediated scenario (SUGRA)~\cite{msugra}, SUSY is
broken in a hidden sector and the breaking is transmitted gravitationally to the observable
sector. This mechanism is elegant, since it only requires already existing fields and
interactions, like gravity. It suffers, however, from the so called SUSY flavour problem,
since it requires a large amount of fine tuning in the squark and slepton mass matriced to
avoid unobserved large flavour-changing neutral current effects.

To cope with the SUSY flavour problem, different SUSY breaking mechanisms have been proposed.
In the Gauge Mediated SUSY Breaking scenario (GMSB)~\cite{gmsb-the} the breaking is transmitted
via gauge forces. This model predicts a very characteristic mass spectrum, with a light gravitino
as the lightest supersymmetric particle (LSP), and typically long-lived next-to-lightest
supersymmetric particles (NLSP).

Anomaly Mediated Supersymmetry Breaking (AMSB)~\cite{branes,amsb} is another interesting solution
to the flavour problem of mSUGRA. Rescaling anomalies in the supergravity Lagrangian always give
rise to soft mass parameters in the observable sector. It follows that anomalies contribute to
SUSY breaking in any case, irrespective of the main symmetry breaking mechanism. We shall refer
to AMSB as the model in which all other components that mediate the SUSY breaking are suppressed
and the anomaly mediation is the dominant mechanism.

The minimal AMSB is very predictive: all the low energy phenomenology can be derived by adding to
the Standard Model (SM) only two extra parameters and one sign. Unfortunately, the minimal AMSB 
model would imply negative squared masses (tachyons) for the sleptons at the electroweak scale. 
One way of getting rid of tachyons is to suppose additional, non-anomaly, contributions to the
SUSY breaking which can generate a positive contribution to the soft masses squared. There are a
few string-motivated solutions that generate such a positive contribution without spoiling the 
renormalization group (RG) invariance of the soft terms. In most cases, such a contribution is
universal for all sfermion masses and, in practice, it is enough to add just one extra parameter
to the model. This arises, for instance, when the visible and the hidden sectors lie in separate
branes that communicate only through gravity~\cite{branes}. There are other solutions~\cite{dterms}
that lead to flavour dependent mass terms; such possibilities are less predictive, since the
sfermion spectrum depends on more parameters, and they will not be investigated further in this
paper. In the following, the minimal AMSB with a single, flavour independent, sfermion mass 
parameter will be considered, as implemented in version 7.63 of the program ISAJET (see below).
However, the characteristic gaugino spectrum of AMSB is the same even for models without such an
universal sfermion mass term, and most of the considerations that follow can be applied also
to them.

The paper is organized as follows. In section~\ref{par:phenomenology} the phenomenology of AMSB
relevant to the search at LEP is shortly reviewed. Section~\ref{par:samples} lists the data and
the event generators used to simulate the signal. Section~\ref{par:searches} describes the
results of the searches for the AMSB signatures in DELPHI. In some cases, searches already
performed in DELPHI were just reinterpreted in the context of AMSB. The descriptions of those
searches can be found in the relevant papers cited in that section. In other cases, which are
described here in more detail, it was necessary to adapt the original techniques to the 
requirements specific to the AMSB scenarios. With no evidence of excesses above the SM
expectations, in section~\ref{par:constraints} the results of the searches are combined
to constrain the parameters of the model and the spectrum of some SUSY particles.

\section{Phenomenology of AMSB}
\label{par:phenomenology}

If there is only one common squared mass term for all scalars, all masses and couplings can be 
derived in terms of just three parameters and one sign:
\begin{itemize}
 \item  the mass of the gravitino, \mhf;
 \item  the ratio of the vacuum expectation values of the Higgs fields, \tanb;
 \item  the common scalar mass parameter \mzero;
 \item  the sign of the Higgs term, \sgnmu.
\end{itemize}
In this context, \mzero\ can even be considered as a phenomenological term that parameterizes the
lack of knowledge of the method with which the sleptons acquire physical masses. 

Low-energy gaugino masses ($M_\lambda$), scalar masses ($M_{\tilde Q}$),and trilinear couplings
($A_y$) in AMSB are given by:
\begin{eqnarray}
   M_\lambda & = & \frac{\beta_g}{g} \, \mhf                                              \\
   M_{\tilde Q}^2 & = & - \frac{1}{4} \left( \frac{\partial\gamma}{\partial g} \beta_g +
              \frac{\partial\gamma}{\partial y} \beta_y \right) \mhf^2 + \mzero^2         \\
   A_y & = &  -\frac{\beta_y}{y} \, \mhf 
\end{eqnarray}
where $g$ are the gauge couplings, $y$ the Yukawa couplings and $\gamma$ and $\beta$ are RG
functions. This soft mass spectrum has distinctive features~\cite{amsb} which can differ from
the usual SUGRA or GMSB scenarios. 

\begin{itemize}

\item The gravitino is heavy (this has several advantages for cosmology~\cite{amsb}).

\item The ratios of gaugino masses at the electroweak scale are determined by the ratios of 
the corresponding $\beta$ functions. Therefore, they assume in a natural way different values 
with respect to the theories with gaugino mass unification at a Grand Unification (GUT) Scale:
\begin{equation}
     M_1 : M_2 : M_3 \,\,\, \simeq  \,\,\,  2.8 : 1 : -8
\end{equation}
These ratios have been computed by including the largest next-to-leading corrections~\cite{amsb}.
Small deviations from these ratios can occur as a result of varying the parameters of the model.
Typical values of $\mu$ allowed by the model imply $M_2<M_1<|\mu|$. As a consequence, the
chargino ($\XPM{i,i=1,2}$) and neutralino ($\XN{j,j=1,4}$) mass eigenstates are rather well
approximated by either pure gaugino or pure higgsino states, with 
$\MXN{1} \sim \MXC{1} \sim M_2$, $\MXN{2} \sim M_1$, $\MXN{{3,4}} \sim \MXC{2} \sim |\mu|$.
Therefore, the lightest chargino and neutralino are always a nearly mass-degenerate doublet of
gauginos, with nevertheless $\MXN{1} < \MXC{1}$; the second lightest neutralino is a gaugino of
intermediate mass; and the heaviest chargino and neutralinos are heavy and higgsino-like.

\item Squark masses are rather insensitive to \mzero. AMSB implies squarks and gluinos much
heavier than the LSP, and completely out of reach at LEP.

\item In the slepton sector, if both the right and the left chiral states receive the same 
$\mzero^2$ contribution, the diagonal entries of the mass matrix are accidentally highly
degenerate. Nearly mass-degenerate and highly mixed same flavour sleptons are a distinctive
feature of the minimal AMSB with a flavour-independent \mzero. The lightest stau is always the
lightest charged slepton. The sneutrinos can be lighter than all charged sleptons, and typically
the stau sneutrino is the lightest sneutrino.

\item The CP-odd neutral Higgs, $A$, is usually much heavier than the $Z$, and the lightest
CP-even neutral Higgs, \hn\, is analogous to the SM one \cite{higgs}. Also the mass of the \hn\
is still more tightly bound than in the usual SUSY scenarios, since it should lie below
120~\GeVcc~\cite{higgs}. Therefore, the lower limit obtained at LEP for the SM Higgs mass already
strongly constrains the AMSB parameter space. Moreover, if such a light Higgs is not to be found
at the Tevatron or, later, at the LHC, the AMSB model itself will be completely ruled out.

\end{itemize}

In the model considered here, only the slepton mass spectrum and, to some extent, the Higgs
depend on the assumptions of a common scalar term \mzero. All other features are characteristic
of any AMSB scenario, independently of the procedure used to cope with the tachyonic masses of
the sleptons.

Since \mzero\ is a free parameter, according to its value there are three possible candidates
for the LSP: the nearly mass-degenerate \XN{1}/\XPM{1}, the \snu\ (for relatively small values
of \tanb\ and \mhf) or the \stau. Scenarios with any of the above as LSP are explored in the
following.

\section{Data and simulation samples}
\label{par:samples}

The results listed in this paper come from searches performed in the DELPHI experiment~\cite{delphi}
at the electron-positron collider LEP of CERN, and interpreted in the context of AMSB. Some of these
searches were originally prepared for different analyses and used unmodified here. Others, were
instead optimized to search specifically for the AMSB signatures. If not otherwise specified in
the text, the reader should refer to the papers cited for the description of the samples of the
data and of the SM background simulation used in the different analyses. 

DELPHI collected a total of approximately 116~pb$^{-1}$ while running at the $Z^0$ pole in
the years from 1989 to 1995 (LEP1). About 694~pb$^{-1}$ of integrated luminosity were harvested
in the LEP2 phase, with centre-of-mass energies ranging from 130 to 208~GeV.

SUSYGEN~\cite{susygen} was used for the simulation of the signal. As it does not allow for the
calculation of the particle spectrum of the AMSB models, the input parameters were set so as to
correspond to a spectrum close to the one resulting from the precise calculations in the AMSB
framework of~\cite{amsb}.

ISAJET \cite{isajet}, since version 7.47, allows the calculation of the particle masses and decay
branching modes of the AMSB model of \cite{branes,amsb} as a function of  the four parameters 
\mzero, \mhf, \tanb\ and \sgnmu. To constrain the allowed space of the parameters, the result of
the searches were compared with the prediction of mass spectra, cross-sections and decay modes as
given by ISAJET~7.63~\footnote{In the code used for the scan some of the later corrections were
applied by hand, as implemented in the subsequent versions of the program~\cite{baer}}.
In that version of ISAJET only one loop contributions are considered in the Higgs sector, while
all two loops terms are included for the running of the gauge couplings. The program was run with
the following SM parameters in input: $\alpha_s(\MZ)=0.118$ and the mass of the top quark at the
mean value of \cite{pdg}, i.e. $m_t=174.3$~\GeVcc. Since $m_t$ is relevant in the definition of
the Higgs mass spectrum, also samples with $m_t=169.2$ and $m _t=179.4$~\GeVcc\ were generated,
which corresponds to plus and minus one standard deviation of the value of \cite{pdg}.

\section{Searches used to investigate the AMSB scenario}
\label{par:searches}

In this section, the searches for topologies predicted by the AMSB model at LEP are reviewed.

\subsection{LEP1 limits}
\label{par:lep1}

The precise measurement of the $Z^0$ width at LEP1~\cite{pdg} was used to place severe constraints
on all possible non-SM contributions. Given the good agreement between the measured total width
and the one predicted by the SM, non-standard contributions are expected to be smaller than
3.2~\MeVcc\ at the 95\%\ confidence level (CL). In particular, this rules out charginos with mass
smaller than 45~\GeVcc at the same CL, independently of their field composition and decay modes.
Such lower bound on the mass of the chargino is not going to be affected even if the more
conservative method of~\cite{moenig} is used to fit the amount of allowed non-SM $Z$ width.
The upper limit on the non-SM invisible width is more model independent, and evaluated to be
2.0~\MeVcc\ at the 95\%~CL~\cite{pdg}. Sneutrinos with mass below 43~\GeVcc\ are incompatible
with that limit. Limits for other sparticles depend both on mass and couplings.

\subsection{Search for nearly mass-degenerate chargino-neutralino}
\label{par:degenerate}

One of the key features of AMSB is the very small difference between the masses of the lightest 
chargino and neutralino. Therefore, the results of the search for nearly mass-degenerate 
chargino and neutralino \cite{susana} can be used to investigate AMSB. 

When the masses of the lightest chargino and neutralino are very close, the visible products of
the decay $\XPM{1}\to\XN{1} ff'$ carry little momentum. Therefore, they are both difficult to 
select and trigger on, and they can become almost indistinguishable from the huge background of
two-photon events at LEP2. Dedicated techniques were used for this search. If there is a 
sneutrino lighter than the chargino two-body leptonic decay modes dominate: this case is treated
in section~\ref{par:chargino}.

For $\DM = \MXC{1}-\MXN{1}$ below approximately 200~\MeVcc, the phase space available for the
decay is limited, and the lifetime can be so long that the chargino produced at the interaction
point is seen as either a heavy stable charged particle in the detector, or as a kink in the
reconstructed track. Long-lived charginos are searched for in DELPHI as single tracks with no
signal (veto) in the gas or liquid radiator of the Cherenkov counter, and/or with an anomalously
high ionization loss in the Time Projection Chamber (TPC). Kinks with both the mother chargino and
the daughter charged decay particle reconstructed in the tracking devices were also searched for.

For \DM\ larger than about 200~\MeVcc, or even less if there are light sneutrinos which increase
the leptonic decay width of the chargino, the lifetime tags are no longer effective. It was however
observed that the signature of a photon with high transverse momentum radiated from the initial 
state (ISR), together with the few soft particles from the decay of the chargino, improves both 
the trigger efficiency of the signal and the rejection of the two-photon background.

Nearly mass-degenerate chargino and neutralino are possible in SUSY only if $M_2\gg |\mu|$, that 
is \XPM{1}\ and \XN{1}\ are both almost pure higgsinos, but this case does not concern AMSB, or if 
$M_2\ll |\mu|$, that is \XPM{1}\ and \XN{1}\ are both almost pure gauginos. To maximize the 
sensitivity to AMSB scenarios, the analysis of~\cite{susana} was redone taking further into account
additional scenarios with light sneutrinos. The search was done under the following hypotheses: 
heavy sneutrinos, that is $\msnu\ge 500$~\GeVcc; 100~\GeVcc\ sneutrinos; sneutrinos with mass 
between $\MXC{1} + 1$~\GeVcc\ and 100~\GeVcc; sneutrinos lighter than $\MXC{1} + 1$~\GeVcc. 
In the last case, where all charginos decay promptly, a stricter requirement on the extrapolation 
to the main event vertex of the charged particle tracks was required: the event was accepted only
if at least two charged particles in it were compatible with coming from the primary \ee\ 
interaction vertex.

With respect to the scenario explored in \cite{susana}, if there is a light sneutrino, either
lighter than the chargino or not more than a couple of \GeVcc\ heavier, the leptonic width gets 
strongly enhanced, and the lifetime shortens. In that case, the efficiencies at the smallest \DM\
explored with the ISR tag, turned out to be larger than the ones computed in \cite{susana} for the
same \DM. On the other hand, as the lifetime shortens, the searches that explicitly rely on it
(stable particles and kinks) lose efficiency.

\begin{figure}[tbhp]
\centerline{
\epsfxsize=18.0cm\epsffile{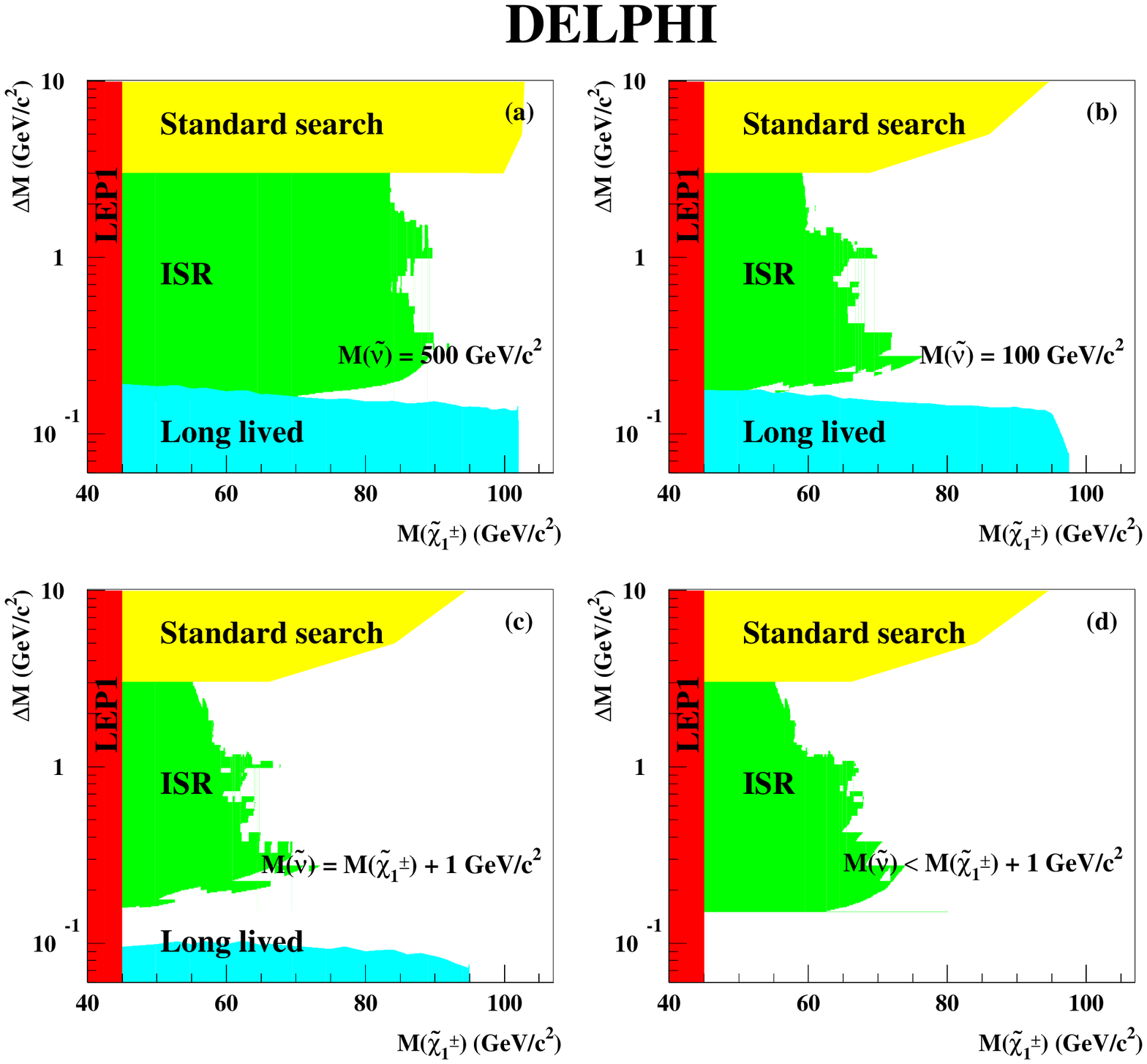} }
\caption[]{ Regions in the plane $(\MXC{1},\DM=\MXC{1}-\MXN{1})$ excluded by DELPHI at
  the $95\%$~CL when the chargino is gaugino-like, as in AMSB. The standard search for 
  high $\DM$ charginos, the search for soft particles accompanied by ISR, and the search
  for long-lived charginos were used. The scenarios constrained in the four plots are:
  (a) $\msnu \ge 500$~\GeVcc;
  (b) $\msnu \ge 100$~\GeVcc;
  (c) $\msnu \ge \MXC{1}+1$~\GeVcc\ (long-lived charginos); 
  (d) $\msnu  <  \MXC{1}+1$~\GeVcc\ (short-lived charginos).
  The exclusions in (a), (b) and (c) hold conservatively also for heavier sneutrinos.
  Charginos lighter that 45~\GeVcc\ were already excluded at LEP1.
       }
\label{fig:limdege}
\end{figure}

Since there was no evidence of an excess in the number of events observed above the SM expectations,
regions in the plane (\MXC{1},\DM) could be excluded at the 95\%\ CL. Figure \ref{fig:limdege} 
shows the regions excluded by the different techniques used in the search for degenerate charginos.
Figure \ref{fig:limdege}~(a) is the same plot with the gaugino exclusion as in \cite{susana}, and 
includes AMSB scenarios when $\msnu \ge 500$~\GeVcc. In figure \ref{fig:limdege}~(b) the exclusion 
was computed for $\msnu=100$~\GeVcc, and therefore it gives conservative predictions in case of 
heavier sneutrinos. Figure \ref{fig:limdege}~(c) was obtained with the minimal chargino 
cross-section (with respect to \msnu) and with the lifetime corresponding to 
$\msnu=\MXC{1}+1$~\GeVcc. This exclusion is conservative for all AMSB scenarios with 
$\MXC{1} + 1~\GeVcc < \msnu < 100~\GeVcc$, since as \msnu\ increases the $s$-$t$ channels 
interference weakens and the cross-section gets larger; moreover, also the lifetime increases, 
thus improving the sensitivity of all searches for long-lived  charginos.
Finally, figure \ref{fig:limdege}~(d) was computed using the minimal chargino cross-section (again
with respect to \msnu) and for short lived charginos. It can be used to constrain AMSB scenarios 
with $\msnu< \MXC{1} + 1$ \GeVcc\ (see also section~\ref{par:chargino}).

To compute these excluded regions, the different channels were combined using the 
multichannel Bayesian method of \cite{obr}. The effect of the systematics uncertainties on the
signal efficiency and on the expected background content was taken into account according to
reference~\cite{appap}.

\subsection{Search for $\XPM{1} \to \snu l^\pm$ decays}
\label{par:chargino}

If the sneutrino is lighter than the chargino, the chargino decays with practically $100\%$
branching ratio into a sneutrino and a charged lepton. Since the upper limits on the chargino
cross-section in the SUGRA scenario were obtained assuming the chargino decaying into 
$\XN{1} W^{*\pm}$~\cite{susana}, those limits cannot be translated directly into limits in the
AMSB scenario. Hence, only the ``leptonic'' search for charginos described in~\cite{susana} was
used to explore the region with $\DSNU= \MXC{1}-\msnu$ larger than 3~\GeVcc.

The analysis selected events with low charged multiplicity and without any reconstructed isolated
photon: events were discarded if they had more than five reconstructed charged particles and if
there was a photon with more than 5~GeV, and isolated by more than 15 degrees from any other
charged or neutral particle. After a preselection obtained with sequential cuts, the final
selection was performed using likelihood ratio functions~\cite{ander} $\lik(\{ x_i\})$ built as
follows: for a set of variables $\left\{ x_{i}\right\}$ (e.g. multiplicities, visible energy, 
acoplanarity~\footnote{Acoplanarity is defined as the complement to $180^\circ$ of the difference
in the azimuthal angle of the two charged particles, or of the two forced jets in case of larger
multiplicity.}, total transverse momentum, fraction of energy in the forward cone, etc.), 
the probability distribution functions of these variables were estimated by normalized 
frequency distributions for the signal (with a \XN{1}\ LSP) and the background samples.
These probability distribution functions were denoted $f_{i}^{S}(x_{i})$ for the signal,
and $f_{i}^{B}(x_{i})$ for the background events that passed the same selection criteria.
Six likelihood ratio functions, one per \DSNU\ region defined as in table~\ref{tab:thomas},
were defined as
\begin{equation}
 \lik = \prod\limits_{i=1}^{n}\frac{f_{i}^{S}(x_{i})}{f_{i}^{B}(x_{i})} \,\,\, .
\end{equation}
Events with $\lik > \likcut$ were selected as candidate signal events. The choice of variables
and the value of \likcut\ were optimized using samples of simulated events, by minimizing the
signal cross-section that was expected to be excluded at 95\%\ CL in the absence of a signal.
This procedure was repeated for every investigated centre-of-mass energy. Basically after the
final selection, the remaining set of events consisted of low-multiplicity events with high
acoplanarity and high missing energy.

Table \ref{tab:thomas} summarizes the number of events observed and expected, and the luminosities
used at the different centre-of-mass energies. The data collected during the year 2000 with and
without the TPC fully operational (see~\cite{susana}) were treated as different channels in the
analyses.

\begin{table}[bt]
\small{
\begin{center}
\begin{tabular}{|c||@{}c@{}|@{}c@{}|@{}c@{}|@{}c@{}|@{}c@{}|@{}c@{}|@{}c@{}|@{}c@{}|} 
\hline
  &  &  &  &  &  &  &  & \\
 \makebox[6.0em]{\small $< \!E_{\mathrm cm}\!>$ (GeV)}
                      & \makebox[5.0em]{191.6} 
                      & \makebox[5.0em]{195.6} 
                      & \makebox[5.0em]{199.6} 
                      & \makebox[5.0em]{201.7} 
                      & \makebox[5.0em]{204.9} 
                      & \makebox[5.0em]{206.7} 
                      & \makebox[5.0em]{208.2} 
                      & \makebox[5.0em]{207.0~}  \\ 
                      &       &        &       &       &       &       &       &        \\
\hline
 {\small $\int \!\! \cal L$ (pb$^{-1}$)}
                      &  25.8 &  76.8  &  84.3 &  40.5 &  78.3 &  78.8 &   7.2 & 60.2   \\ 

\hline \hline
 & \multicolumn{8}{|c|}{3 $\leq$ \DSNU\ $<$ 5~\GeVcc  } \\ \hline
{\small Data}  & 2 & 13 & 17 & 7 &  8 & 11 &  1  & 10 \\
{\small MC}    & 6.0 $^{+ 0.7} _{- 0.4}$ &
                17.4 $^{+ 1.9} _{- 1.2}$ &
                17.9 $^{+ 1.7} _{- 1.2}$ &
                 8.7 $^{+ 0.8} _{- 0.6}$ &
                 9.8 $^{+ 1.3} _{- 0.8}$ &
                 9.9 $^{+ 1.3} _{- 0.8}$ &
                 0.9 $^{+ 0.1} _{- 0.1}$ &
                18.8 $^{+ 1.5} _{- 1.1}$                           \\
\hline
 & \multicolumn{8}{|c|}{ 5 $\leq$ \DSNU\ $<$ 10~\GeVcc } \\ \hline
{\small Data}  & 2 & 2 &  4 &  5 &  0 &  0 &  0  &  4 \\
{\small MC}    & 1.3 $^{+ 0.4} _{- 0.2}$ &
                 3.8 $^{+ 1.1} _{- 0.5}$ &
                 4.2 $^{+ 1.0} _{- 0.4}$ & 
                 2.1 $^{+ 0.5} _{- 0.2}$ &
                 1.0 $^{+ 0.7} _{- 0.2}$ &
                 1.0 $^{+ 0.7} _{- 0.2}$ &
                 0.1 $^{+ 0.1} _{- 0.1}$ &
                 3.6 $^{+ 0.8} _{- 0.4}$                            \\
\hline
 & \multicolumn{8}{|c|}{ 10 $\leq$ \DSNU\ $<$ 25~\GeVcc } \\ \hline
{\small Data}  & 1 & 5 &  7 &  1 &  5 &  3 &  0 &  3  \\
{\small MC}    & 1.6 $^{+ 0.4} _{- 0.1}$ &
                 5.0 $^{+ 1.0} _{- 0.3}$ & 
                 5.1 $^{+ 0.9} _{- 0.3}$ &
                 2.5 $^{+ 0.5} _{- 0.2}$ &
                 3.7 $^{+ 0.8} _{- 0.3}$ &
                 3.7 $^{+ 0.8} _{- 0.3}$ &
                 0.3 $^{+ 0.1} _{- 0.1}$ &
                 2.3 $^{+ 0.6} _{- 0.2}$                            \\

\hline
 & \multicolumn{8}{|c|}{ 25 $\leq$ \DSNU\ $<$ 35~\GeVcc } \\ \hline
{\small Data}  & 2 &11 &  5 &  3 &  5 &  8 &  0 &  3 \\
{\small MC}    & 2.8 $^{+ 0.4} _{- 0.2}$ &
                 9.0 $^{+ 1.1} _{- 0.4}$ & 
                 8.5 $^{+ 0.9} _{- 0.3}$ &
                 4.1 $^{+ 0.5} _{- 0.2}$ &
                 5.5 $^{+ 0.8} _{- 0.3}$ &
                 5.6 $^{+ 0.8} _{- 0.3}$ &
                 0.5 $^{+ 0.1} _{- 0.1}$ &
                 5.3 $^{+ 0.7} _{- 0.2}$                            \\
\hline
 & \multicolumn{8}{|c|}{ 35 $\leq$ \DSNU\ $<$ 50~\GeVcc } \\ \hline
{\small Data}  & 6 &20 & 10 &  4 & 11 & 10 &  2 & 10 \\
{\small MC}    & 5.5 $^{+ 0.4} _{- 0.2}$ &
                16.0 $^{+ 1.1} _{- 0.5}$ & 
                15.5 $^{+ 0.9} _{- 0.4}$ &
                 7.3 $^{+ 0.5} _{- 0.2}$ &
                12.8 $^{+ 0.8} _{- 0.3}$ &
                12.9 $^{+ 0.8} _{- 0.3}$ &
                 1.2 $^{+ 0.1} _{- 0.1}$ &
                11.7 $^{+ 0.7} _{- 0.3}$                            \\
\hline
 & \multicolumn{8}{|c|}{ 50~\GeVcc\ $\leq$ \DSNU } \\ \hline
{\small Data}  & 9 &32 & 14 &  5 & 22 & 18 &  0 & 16 \\
{\small MC}    & 8.4 $^{+ 0.5} _{- 0.2}$ &
                23.8 $^{+ 1.2} _{- 0.6}$ & 
                24.0 $^{+ 1.1} _{- 0.5}$ &
                11.5 $^{+ 0.6} _{- 0.3}$ &
                18.6 $^{+ 0.8} _{- 0.3}$ &
                18.7 $^{+ 0.8} _{- 0.3}$ &
                 1.7 $^{+ 0.1} _{- 0.1}$ &
                14.2 $^{+ 0.6} _{- 0.3}$                            \\
\hline
 & \multicolumn{8}{|c|}{ TOTAL (logical .OR. between different \DSNU\ windows)} \\ \hline
{\small Data} & 10 &52 & 34 & 13 & 37 & 36 &  3 & 31 \\
{\small MC}    & 15.7 $^{+ 0.8} _{- 0.5}$ &
                 46.2 $^{+ 2.2} _{- 1.6}$ & 
                 43.9 $^{+ 1.9} _{- 1.3}$ &
                 21.2 $^{+ 0.9} _{- 0.7}$ &
                 33.2 $^{+ 1.5} _{- 1.0}$ &
                 33.4 $^{+ 1.5} _{- 0.9}$ &
                  3.1 $^{+ 0.1} _{- 0.1}$ &
                 38.8 $^{+ 1.7} _{- 1.2}$                            \\
\hline
\end{tabular}
\end{center}
            }
\caption[.]{
\label{tab:thomas}
The number of events observed in data and the expected number of background events in the search
for a pair of charginos both decaying into a sneutrino and a charged lepton, at the centre-of-mass
energies collected by DELPHI in 1999 and 2000. The last column corresponds to the data collected in
the year 2000 with the TPC not fully operational (at their mean centre-of-mass energy).}
\end{table}

\begin{figure}[ht]
\begin{center}
\begin{tabular}{cc}
\hspace*{-1.3cm}
\begin{minipage}[c]{7.5cm}
\epsfxsize=9.3cm  
\epsffile{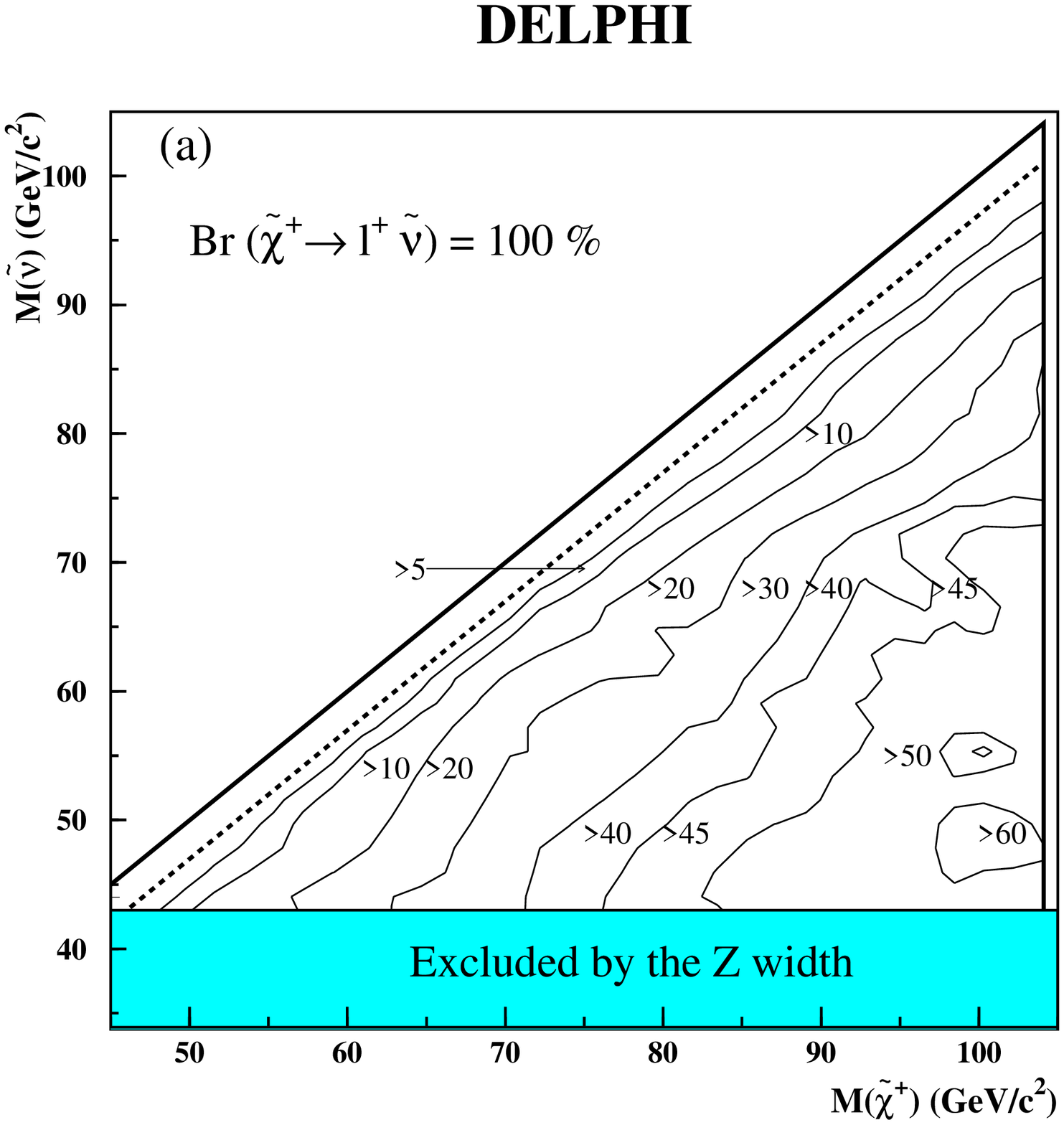}
\end{minipage}
&
\hspace*{0.4cm}
\begin{minipage}[c]{7.5cm}
\epsfxsize=9.3cm  
\epsffile{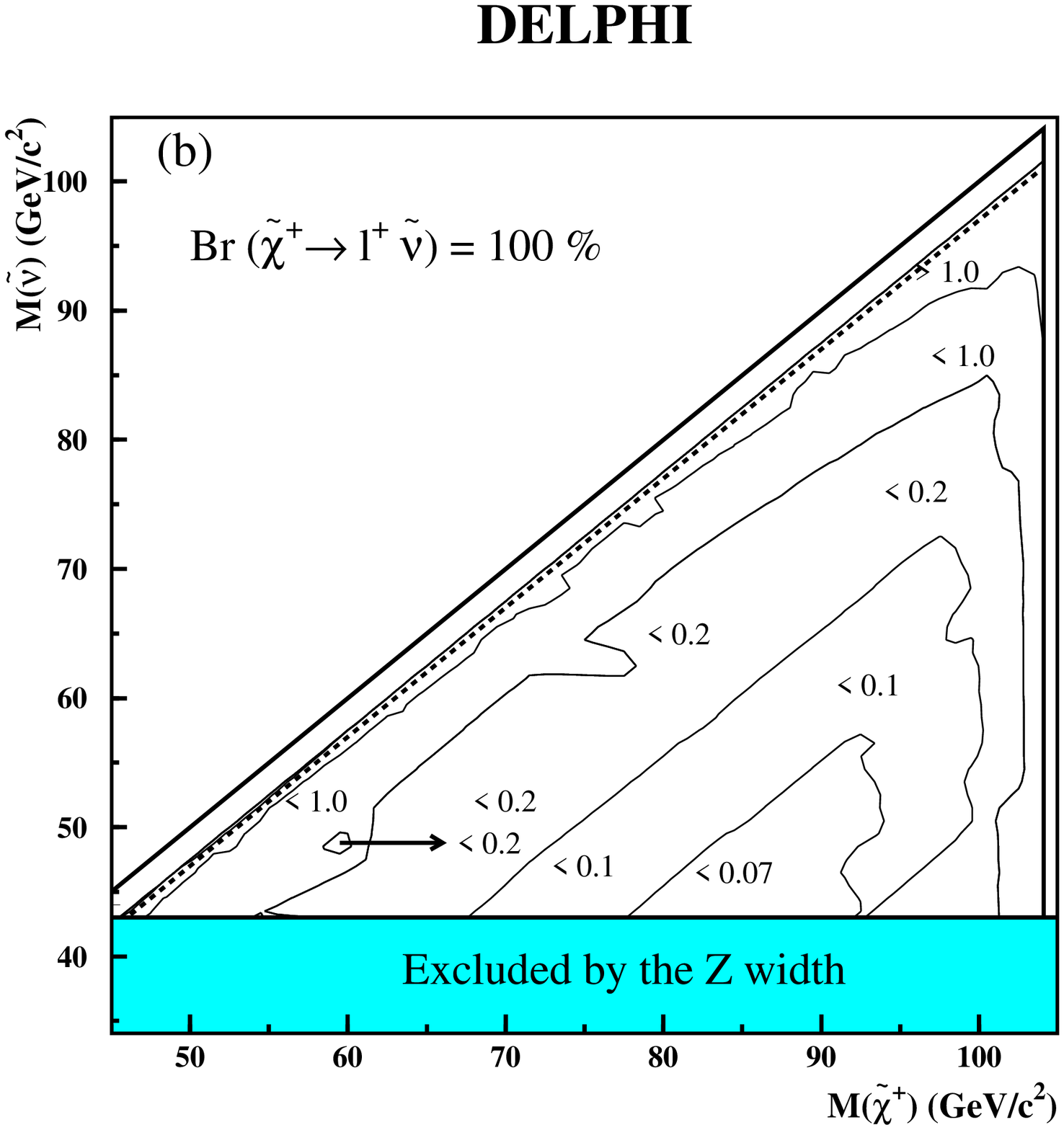}
\end{minipage}
\\
\end{tabular}
\vspace{0.5cm}\\
\begin{tabular}{cc}
\hspace*{-0.9cm}
\begin{minipage}[c]{7.5cm}
\hspace*{-0.5cm}
\epsfxsize=9.3cm  
\epsffile{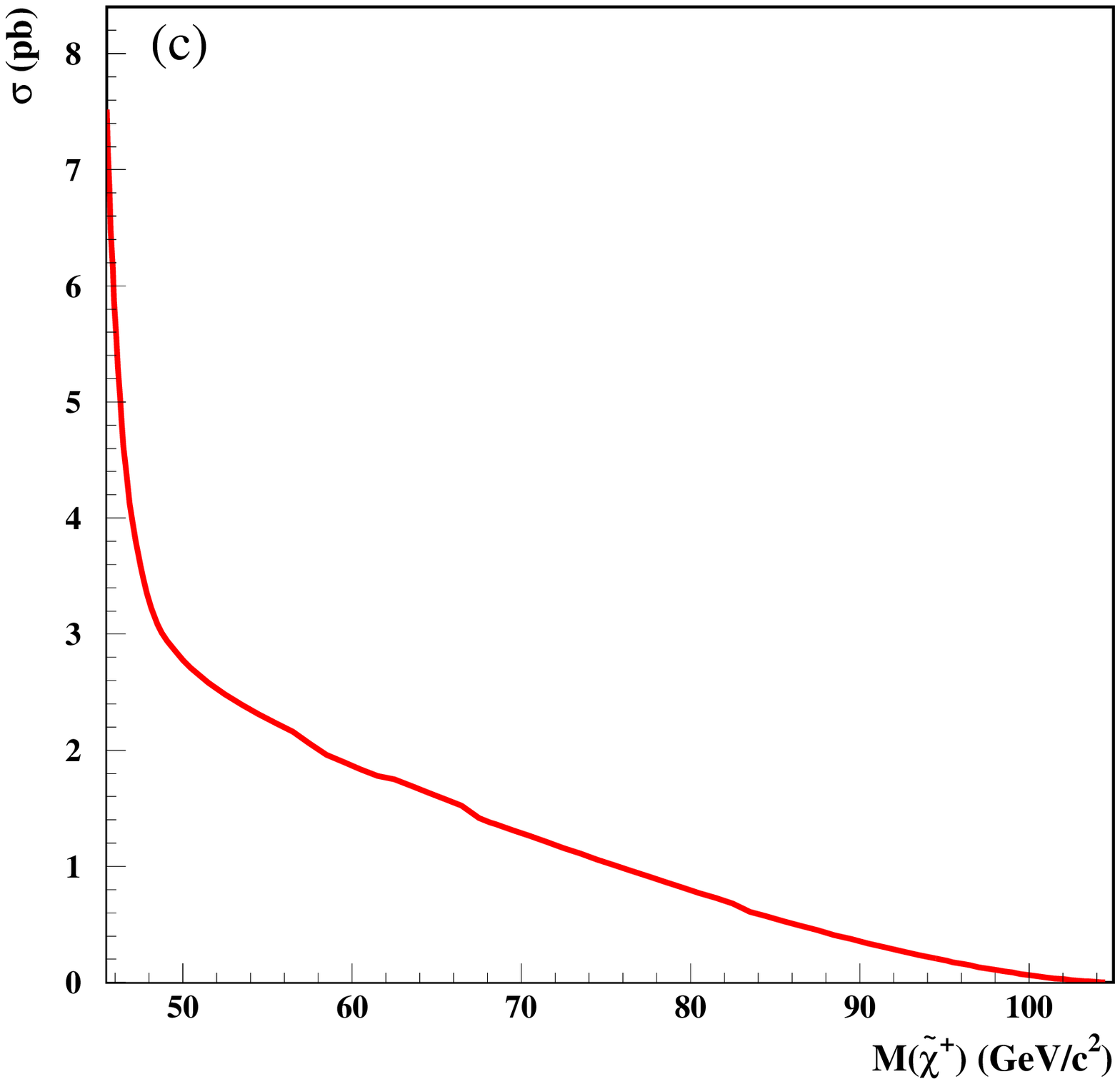}
\end{minipage}
&
\hspace*{0.01cm}
\begin{minipage}[c]{7.5cm}
\epsfxsize=9.3cm  
\epsffile{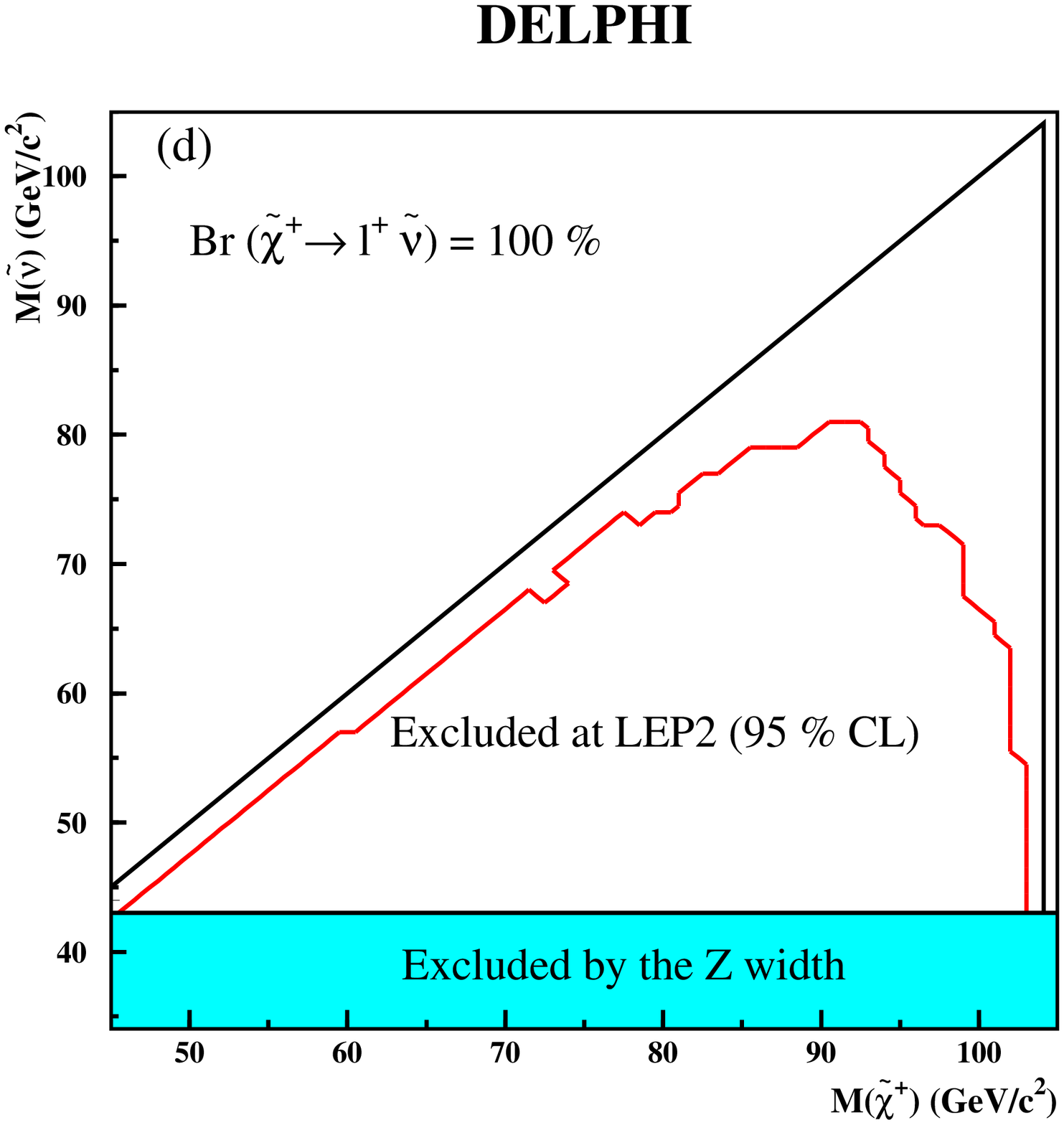}
\end{minipage}
\\
\end{tabular}
\vspace{0.5cm}
\caption[]{ (a) Chargino pair production detection efficiencies~(\%) for the fully leptonic decay
    channel at \rs=208.2~GeV in the (\MXC{1},\msnu) plane; a 100\% branching ratio of
    $\XPM{1}\to\snu l^{\pm}$ is assumed.
    (b) Equivalent excluded cross-section at the 95\%~CL (in pb) at 208.2 GeV.
    (c) Minimal expected $\eeto\XP{1}\XM{1}$ cross-section in AMSB, as function of the mass of
    the chargino.
    (d) Region excluded at the 95\%~CL in the plane (\MXC{1},\msnu) by the search described in the
    text. Sneutrinos lighter than 43~\GeVcc\ were already excluded at LEP1.
    The dotted lines in figures (a) and (b) bound the range of $\DM=\MXC{1}-\msnu$ searched for
    by this channel.
           } 
\label{fig:thomas}
\end{center}
\end{figure}

The efficiencies at the centre-of-mass energy of 208 GeV of the fully leptonic selection are 
plotted in figure~\ref{fig:thomas}~(a) as function of the chargino and sneutrino masses. Since up
to five visible charged particles were allowed and no leptonic identification was required, those 
efficiencies had only very little dependence on the flavour of the charged lepton in the final 
state. The efficiencies of figure \ref{fig:thomas}~(a) were computed using events simulated with
three body $\XP{1}\to l^+\nu \XN{1}$ decays. It was however verified with samples of fully 
simulated events with the two body decay $\XP{1}\to l^+\snu$ searched for in AMSB, that the 
efficiencies used in the analysis were never larger than the ones expected for the two body decays.
This was expected, since the momentum of the visible charged particles is on average larger in the
two body sample than in the three body one. In particular, the efficiencies used are fully 
compatible with the true ones at large \DSNU, and they become up to one and a half times smaller
when \DSNU\ approaches 3~\GeVcc.
Therefore, the limits obtained are never overestimated, and possibly conservative at small \DSNU.

No significant excess was observed above the SM expectations. After having combined all channels
with the multichannel Bayesian method of \cite{obr}, figure \ref{fig:thomas}~(b) displays the 
95\%\ CL upper limit of the chargino cross-section at the reference centre-of-mass energy of 
208~GeV, as function of the masses of the chargino and of the sneutrino, and assuming 
$\mathrm{BR}(\XPM{1}\to\snu l^\pm)=1$. If that exclusion is compared with the theoretical
expectation of the same cross-section (in figure \ref{fig:thomas}~(c) the minimal expected
$\eeto \XP{1}\XM{1}$ cross-section is shown as function of \MXC{1}), a region in the plane
$(\MXC{1},\msnu)$ can be excluded at the same confidence level. Such excluded region is shown
in figure~\ref{fig:thomas}~(d).

The exclusion when $0<\DSNU<3$~\GeVcc, as obtained with the results of the search for nearly 
mass-degenerate charginos, can be derived from figure~\ref{fig:limdege}~(d), by simply substituting
\DM\ with \DSNU\ in the ordinate. The same observation on the conservativeness of the limits
obtained when the chargino decays into two bodies also holds for the search at $\DSNU<3$~\GeVcc,
since the efficiencies are estimated from samples of three body decays.
For $\MXC{1}<55$~\GeVcc, only the narrow band $0<\DSNU<200$~\MeVcc\ cannot be excluded.

If also the stau, or some other charged slepton, has a mass which is intermediate between the mass
of the chargino and that of the sneutrino, figure~\ref{fig:thomas}~(b) should be interpreted as the
95\%\ CL upper limit of the chargino cross-section times its branching ratio into $l^{\pm}\snu$.

\subsection{Search for $\eeto \XN{1}\XN{2}$ }
\label{par:neutralino}

Searches for \XN{1}\XN{2} production with $\XN{2} \to q\bar{q} \XN{1}$,
$\XN{2} \to \mu^+\mu^- \XN{1}$, $\XN{2} \to \ee \XN{1}$, $\XN{2} \to \Zn \XN{1}$, and
$\XN{2} \to \stau \tau$ decays have been presented in \cite{susana}. Limits for production
cross-section times branching ratio to the corresponding final state ranged typically from 0.05~pb
to 0.2~pb, depending primarily  on the mass difference $\MXN{2}-\MXN{1}$.

Since in AMSB $M_1/M_2 \sim 2.8$, and $\MXN{1} \sim \MXC{1} \sim M_2$ and $\MXN{2}\sim M_1$, there
is relatively little phase space available for the production of \XN{1}\XN{2} at LEP energies. 
Only if \XN{1} is sufficiently light a \XN{2}, which is almost three times as heavy as the \XN{1},
can be produced in association, as $\MXN{1}+\MXN{2}$ must be below the centre-of-mass energy 
available in the \ee\ collision. In this case, the \XN{2}
decays mainly to $\XN{1} \Zn$ and $\XPM{1} W^\mp$ \cite{isajet}. For the $\XN{2} \to \XN{1} \Zn$
decay, the results of the neutralino searches presented in \cite{susana} can be directly used.
Since in AMSB scenarios the chargino is nearly mass-degenerate with the neutralino, the decay
$\XN{2} \to \XPM{1} W^\mp$,  with $\XPM{1} \to \pi^\pm \XN{1}$ and $W \to q \bar{q}'$, results in
the same visible final state as  $\XN{2} \to \XN{1} q\bar{q}$. Also in this case, the limits on
\XN{1}\XN{2} production with the above final state presented in \cite{susana} can be used. On the
other hand, when the $W$ decays leptonically, the visible objects in the final states are different
from those of the standard search for neutralinos, because of the soft particles from the chargino
decay, which can be relevant in a low-multiplicity environment. A dedicated search would be needed,
but is not considered in the present paper.

If there are sleptons with a mass between \MXN{1} and \MXN{2}, cascade decays of \XN{2} can take
place: $\XN{2} \to \sle l$, $\sle \to \XN{1} l$. In this case there are two mass differences
(\DSLE) characterizing the process: $\MXN{2}-\msle$ and $\msle-\MXN{1}$. 
It was verified that if $\sle = (\smu,\sel)$ the results of the searches for $\eeto\XN{2}\XN{1}$,
where $\XN{2} \to \mu^+\mu^- \XN{1}$ or $\XN{2} \to \ee \XN{1}$, can be used, provided that from
the \DSLE\ definitions above the one giving the more conservative result is used in place of
$\MXN{2}-\MXN{1}$.

\begin{figure}[tbhp]
\centerline{
\epsfxsize=14.0cm\epsffile{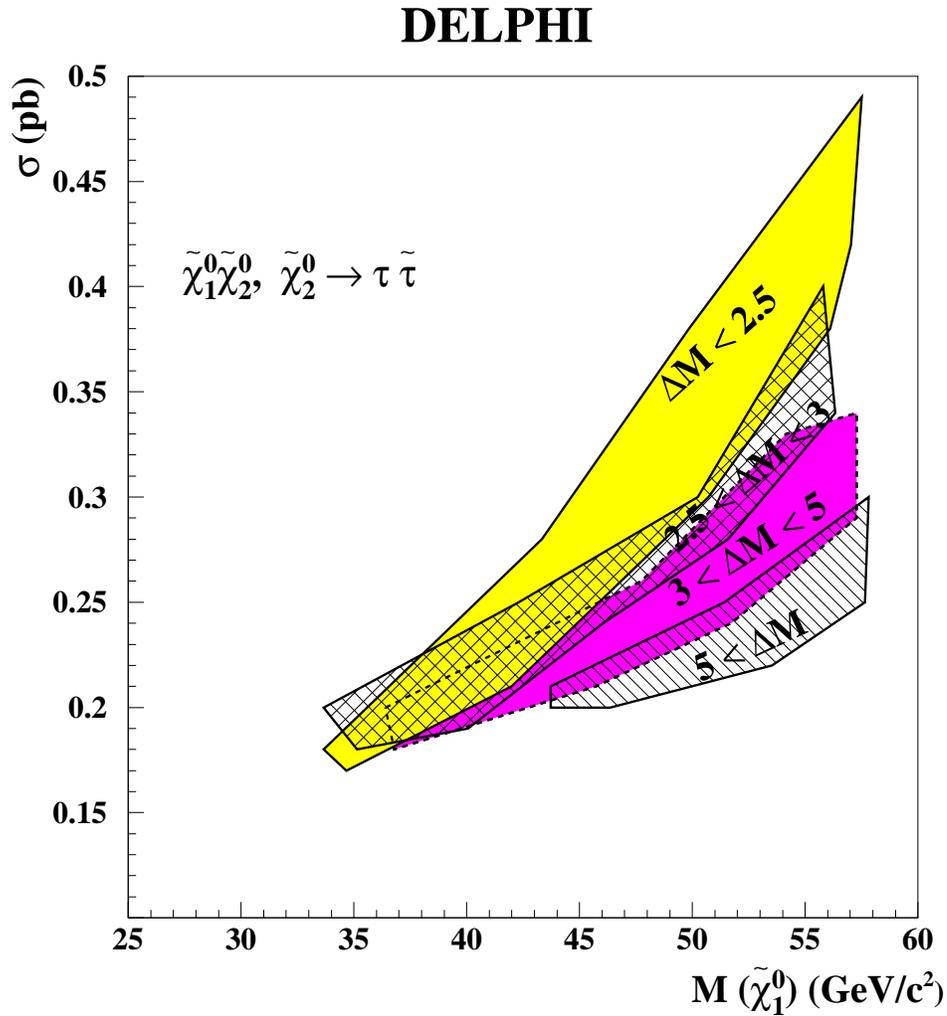} }
\caption[]{ Cross-section limits at the 95\%~CL for the \XN{1}\XN{2} production when \XN{2} decays
entirely to $\stone\tau$. The upper limits are shown for several ranges of \DM=\mstone-\MXN{1} in
\GeVcc. 
The widths of the bands are due to dependence of the limit on \DM\ and to statistical fluctuations
of the efficiency due to limited  Monte Carlo statistics.}
\label{fig:neuttau}
\end{figure}

The lightest stau, \stone, is typically the lightest charged slepton in AMSB. For \stone\ as the
intermediate slepton, the tau cascade search described in \cite{susana} was studied in a wider
range of $\mstone-\MXN{1}$.
The tau cascade search is sensitive to \XN{1}\XN{2} production with $\XN{2} \to \stau\tau$
and $\stau \to \XN{1}\tau$, where the second $\tau$ produced has very low energy.
At the preselection level, well reconstructed low-multiplicity events with missing energy,
missing mass, and no more than two reconstructed jets were selected. In particular, the total
visible energy including badly reconstructed tracks was required to be less than 140~GeV, the
number of charged particles was required to be at least two and at most eight, and the number
of neutral particles to be less than five. Two or more of the charged particles also had to
satisfy stricter criteria on reconstruction and impact parameters. There was no evidence of
excess above the SM expectations after the selection (see~\cite{susana}). The resulting
cross-section upper limits at the 95\%~CL are shown in figure \ref{fig:neuttau}.

Light sneutrinos lead to an undetectable $\XN{2} \to \snu \bar\nu$ and $\snu \to \XN{1}\nu$
decay chain.

\subsection{Search for a charged slepton as the LSP}
\label{par:stable}

In a scan of the parameter space performed with ISAJET~7.63~\cite{isajet} no points were 
obtained whera a charged slepton is lighter than the \XN{1}. However, the calculations
in~\cite{amsb} still allow a small region in the space of the AMSB parameters with the
$\stone$ being the LSP. In this case, if R-parity is conserved, the stau must be stable.
The DELPHI results of the search for heavy stable charged particles were presented in
\cite{gmsb-pap}, together with the description of the method used in the analysis. 

The left and right-handed staus are expected to be almost maximally mixed in AMSB~\cite{amsb}.
Reference~\cite{gmsb-pap} showed that the results of the search for heavy stable charged particles
in DELPHI can exclude a stable $\stone$ with mass below 96~\GeVcc\ at the 95\% CL, even at the
level of mixing that gives the lowest $\stone^+\stone^-$ production cross-section.

\subsection{Search for cascades from sleptons}
\label{par:staucascades}

The decay $\sle^{\pm} \to \XPM{1} \nu_l$ is practically undetectable, due to the softness of the
visible decay products of the chargino. It accounts, however, for two thirds of the total decay
width, if the chargino and the neutralino are the only SUSY particles lighter than the charged
slepton. The only visible cascades originating from that slepton (in particular a stau, 
since it is expected to be the lightest) in AMSB are therefore:
\begin{itemize}
\item $\stone\to\XN{1}\tau$, the same channel searched for in MSSM;
\item $\stone\to\snu_{\tau}ff'$, with visible final states that can be similar to the chargino ones.
\end{itemize}

In the case of sneutrino production, the decay $\snu \to \XN{1} \nu$ is clearly invisible. 
On the other hand, $\snu \to \XP{1} l^-$ is observable using techniques similar to those used
in the usual searches for sleptons~\cite{susana}.

Limits on cross-section times branching ratio can be derived by interpreting the results
of the searches for ``standard'' charginos and sleptons listed in~\cite{susana}. No optimization
by means of a dedicated study of those cascades was attempted for the present paper.

\subsection{Search for the SUSY Higgs boson}
\label{par:higgs}

Since in the range of the AMSB parameters explored in this paper $\MA \gg \MZ$, the lightest
supersymmetric neutral Higgs \hn\ has the same couplings as the SM Higgs boson, and the limits
obtained on the mass of the Higgs in the SM can be translated into the same lower limits on the
mass of the \hn\ in AMSB, provided that the decay branching fractions of the Higgs into
supersymmetric particles are negligible.

If $\MA\gg\MZ$, the \hn\ can be produced at LEP only in association with the \Zn\ (higgsstrahlung), 
and with the same cross-section as in the SM. When there are SUSY particles lighter than $\MSH/2$,
decays of the \hn\ into those particles are allowed. This is the case for AMSB, when there are
light winos, sneutrinos or charged sleptons. Possible SUSY decays of the \hn\ are:
\begin{itemize}
\item $\hn \to \XN{1}\XN{1}$, $\XP{1}\XM{1}$, $\snu\snu$, all invisible or practically invisible
      in AMSB, apart from some possible cascades;
\item $\hn \to \sle^+ \sle^-$, the visibility of which depends on the mass difference between the
      slepton and the LSP.
\end{itemize}

The DELPHI bound on the SM Higgs mass is $M_H>114.1$~\GeVcc\ at the 95\%~CL~\cite{smhiggs}. 
DELPHI measured also the upper limit on the production cross-section of an invisibly decaying
Higgs boson~\cite{invhiggs}. This leads to exclude a Higgs boson with mass below 112.1~\GeVcc,
if it has a 100\% branching ratio into invisible particles. Reference~\cite{invhiggs} shows how
the lower limit on the mass of the lightest supersymmetric Higgs boson depends on the branching
fraction into invisible states, assuming that the production cross-section and all other decay
modes are SM-like. That limit starts from 114.1~\GeVcc\ when $\mathrm{BR}(\hn\to\mathrm{inv.})=0$,
that is when the \hn\ decays as the SM Higgs; it reaches a minimum of 111.8~\GeVcc\ when both
visible and invisible decay modes are present simultaneously; and it goes up again to 112.1~\GeVcc\
when $\mathrm{BR}(\hn\to\mathrm{inv.})=1$. The same limits on \MSH\ apply in AMSB, provided there
are no visible SUSY decays with sizeable branching fractions.

\section{Constraints on the AMSB spectrum}
\label{par:constraints}

The negative results of the searches described in this paper were used to constrain the AMSB
parameter space. To do so, the experimental exclusions measured were compared with the mass
spectra produced by ISAJET~7.63~\cite{isajet}. A scan over the AMSB parameters was carried out by
varying them in the following ranges: $1<\mhf<50$~\TeVcc; $0<\mzero<1000$~\GeVcc; $1.5<\tanb<35$;
both  positive and negative $\mu$. 900,000 SUSY points were generated by choosing at random the
parameters within the bounds above. 500,000 of those points were generated with the mass of the
top quark at 174.3 \GeVcc, the others having been divided between $m_t=169.2$ and $179.4$ \GeVcc,
as explained in section \ref{par:samples}. 
A bigger density of points was allowed in the regions of the space of the parameters where the
expected limits lied as well as in the regions where some structure was observed.

With the AMSB model as implemented in that version of ISAJET, only the negative results from the 
search for nearly mass-degenerate chargino and neutralino, the search for neutral SM-like and 
invisibly decaying Higgs, the search for charginos decaying into a sneutrino and a charged
lepton, and the limit on the non-SM $Z$ width from LEP1 were used to constrain the model 
parameters. The other searches described in this paper were found to provide no additional
constraints to the model. They have been listed all together as well, in order to allow tests
of any deviations from the implementation of AMSB as coded in ISAJET.

\begin{figure}[tbhp]
\centerline{
\epsfxsize=16.0cm\epsffile{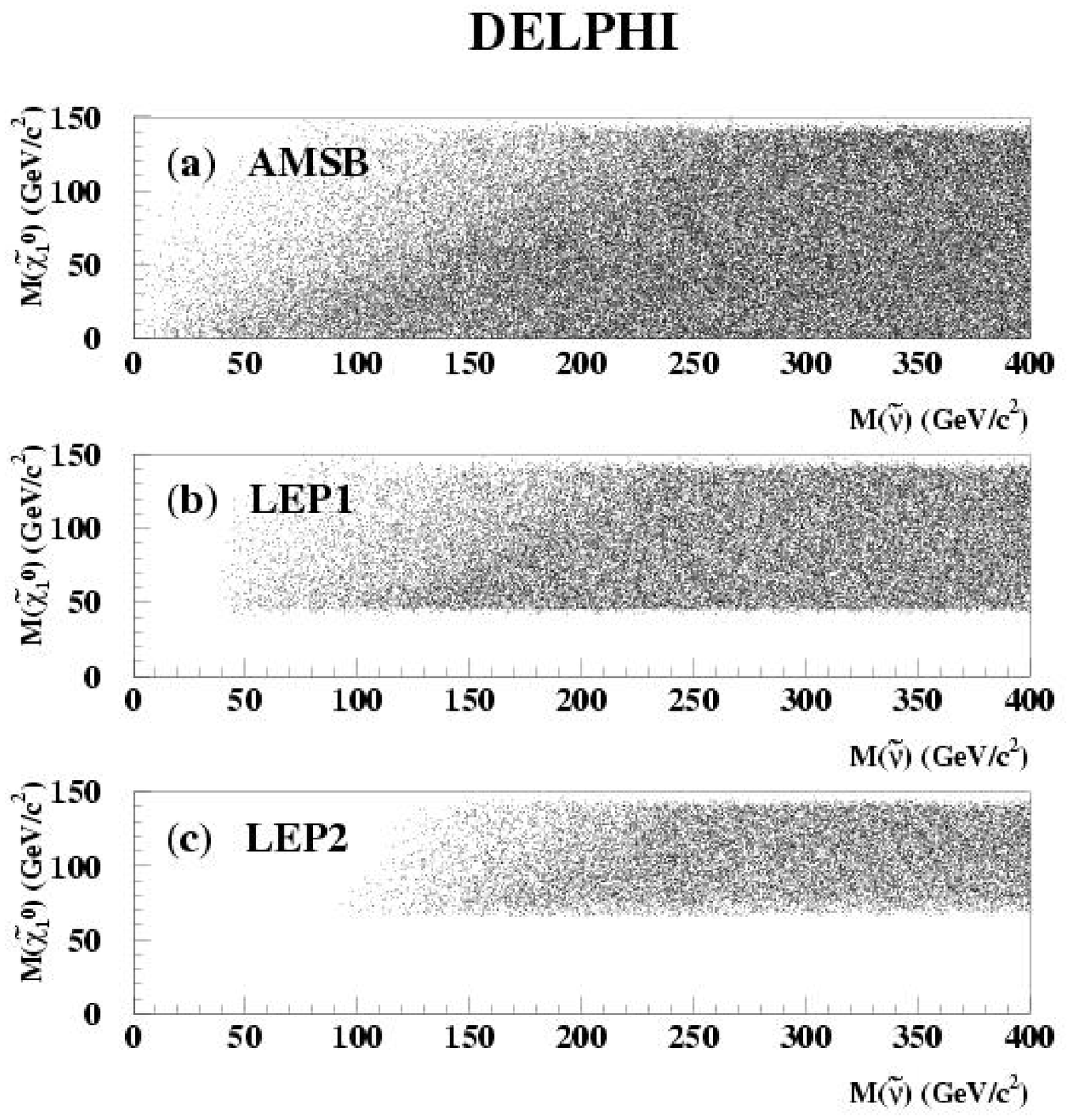} }
\caption[]{ (a) physically allowed \MXN{1} and \msnu\ in AMSB, as obtained in a scan of the
        AMSB parameter space with ISAJET, as described in the text. (b) points remaining after
        applying the chargino and sneutrino mass bounds of LEP1. (c) set of points from
        the scan remaining after considering all the results of the searches described in
        this work. No points survived for which $ \MXN{1}<66$~\GeVcc\ or $\msnu<95$~\GeVcc.
       }
\label{fig:isajet-chisnu}
\end{figure}

The following figures \ref{fig:isajet-mm}, \ref{fig:isajet-hig} and \ref{fig:isajet-chisnu} refere
to the scan done for the central value of $m_t$. Similar figures were obtained in correspondence
of the two other values of $m_t$ considered, and their outcomes are summarized in 
table~\ref{tab:limits}.

Figure \ref{fig:isajet-mm}~(a) shows the points in the plane ($\mzero,\mhf$) generated with ISAJET.
The region of the space with no points was not allowed, because one or more sparticles would be
tachyonic. This implies a certain degree of correlation between \mzero\ and \mhf, since by cutting
away slices at low \mhf\ the value of the lowest admissible \mzero\ increases.
Figure~\ref{fig:isajet-mm}~(b) shows the points that remain after the  application of the
model-independent bounds on the chargino and sneutrino masses obtained at LEP1. 
Finally, in figure~\ref{fig:isajet-mm}~(c) the points that remain after having applied
all the results of the searches described in this paper are displayed.

Since the model prefers a light Higgs, most of the exclusion in the space of the AMSB parameters 
arises from the negative results of the searches for the SM and the invisibly decaying Higgs boson.
The negative results of the other searches enlarged further the rejection, especially at low
\mhf\ (chargino searches) and low \mzero\ (searches with sleptons). The effect of the search for 
the standard and invisible Higgs can be seen in figure~\ref{fig:isajet-hig}. 
Figure~\ref{fig:isajet-hig}~(a) shows all the points generated with ISAJET in the plane 
($\MSH,\tanb$). Figure~\ref{fig:isajet-hig}~(b) shows all the points remaining after the LEP1 
chargino and sneutrino bounds, and the exclusions obtained with the searches for SUSY particles,
but excluding the Higgs, at LEP2. Figure~\ref{fig:isajet-hig}~(c) shows the points surviving after
the negative results of the SM and invisibly decaying Higgs searches in DELPHI. One can notice how
the search for the Higgs boson and the search for the other SUSY particles at LEP are complementary
in excluding certain regions in the space of the AMSB parameters. Figure~\ref{fig:isajet-hig}~(d)
shows that, after applying the full set of results presented in this paper to constrain AMSB, no
point with a mass of the lightest Higgs below the SM limit of 114.1~\GeVcc\ survived.

It is interesting to observe the impact of the searches for AMSB on some sparticle masses. 
Figure \ref{fig:isajet-chisnu} shows the number of points generated by ISAJET and passing the three
steps of selection as in figure \ref{fig:isajet-mm}, as a function of the mass of the lightest
neutralino and of the lightest sneutrino. Neutralinos lighter than 66~\GeVcc\ and sneutrinos 
lighter than 95~\GeVcc\ are excluded in AMSB. 

\begin{table}[bt]
\normalsize{
\begin{center}
\begin{tabular}{|c|c|c|} 
  \cline{2-3} 
 \multicolumn{1}{c|}{\makebox[4.5em]{}}   &  \makebox[8.0em]{}  &  \makebox[8.0em]{}  \\
 \multicolumn{1}{c|}{}            &  $\mu < 0$     &  $\mu > 0 $    \\
 \multicolumn{1}{c|}{}            &                &                \\
 \hline
          &                              &                  \\
 \mzero   &  $>183$ (211, 174)   \GeVcc  &  $> 156$ (181, 147) \GeVcc    \\
 \mhf     &  $>26.3$ (30.0, 24.5) \TeVcc &  $> 23.0$ (26.1, 21.4) \TeVcc \\ 
 \tanb    &  $> 5.7$ (7.0, 4.9)          &  $> 3.8$ (4.6, 3.4)           \\
          &                              &                               \\
 \mchi    &  $>  73$ (83, 67)   \GeVcc   &  $>  66$ (74, 63)  \GeVcc     \\ 
 \msnu    &  $> 114$ (131, 104) \GeVcc   &  $>  95$ (116, 85) \GeVcc     \\ 
 \msle    &  $>  75$ (90, 70)   \GeVcc   &  $>  68$ (78, 66)  \GeVcc     \\ 
          &                              &                               \\
\hline
\end{tabular}
\end{center}
            }
\caption[.]{
\label{tab:limits}
Bounds on the AMSB parameters and on the sparticle masses, as a function of sign($\mu$), obtained
by applying the 95\%~CL limits derived in the searches for AMSB scenarios. Given the small mass
splitting, at the level of few hundred \MeVcc, \mchi\ can be viewed both as \MXN{1} or \MXC{1}.
\msnu\ is the mass of the lightest sneutrino, always the tau sneutrino in the model. \msle\ refers
to the lightest charged slepton, which is always the stau in AMSB. Within parenthesis are listed
the same bounds obtained with $m_t$ respectively below and above one standard deviation, as from
\cite{pdg}, from the central value of 174.3~\GeVcc.}
\end{table}

Table~\ref{tab:limits} summarizes the bounds on the AMSB parameters and on the mass of some 
sparticles obtained by applying the 95\%~CL exclusions from the searches described in the previous
paragraphs to the ISAJET spectra. They are listed separately as function of the sign of $\mu$ and
of the value of the mass of the top quark used in the simulation. Small shifts of those bounds are
still possible, in principle, when applying the full next-to-leading order corrections to the model.


Given the bounds listed in table~\ref{tab:limits}, the possible AMSB explanation for a light
sneutrino ($\msnu \leq 80$~\GeVcc), which was suggested to cure some of the discrepancies in the
fit of precision electroweak data~\cite{ridolfi}, is likely to be ruled out by the results of this
analysis.

\section{Conclusions}

The results of the searches performed using the data collected with the DELPHI detector at LEP, 
and relevant to explore AMSB scenarios, have been presented. An interpretation of the limits 
obtained in searches motivated by other SUSY breaking scenarios was used whenever appropriate.
In addition, some of the searches were developed specifically to improve the sensitivity to AMSB.
There is no evidence for a signal beyond the Standard Model, and limits are set on the sparticle
production in the AMSB framework.

\subsection*{Acknowledgements}
\vskip 3 mm
 We are greatly indebted to our technical 
collaborators, to the members of the CERN-SL Division for the excellent 
performance of the LEP collider, and to the funding agencies for their
support in building and operating the DELPHI detector.\\
We acknowledge in particular the support of \\
Austrian Federal Ministry of Education, Science and Culture,
GZ 616.364/2-III/2a/98, \\
FNRS--FWO, Flanders Institute to encourage scientific and technological 
research in the industry (IWT), Federal Office for Scientific, Technical
and Cultural affairs (OSTC), Belgium,  \\
FINEP, CNPq, CAPES, FUJB and FAPERJ, Brazil, \\
Czech Ministry of Industry and Trade, GA CR 202/99/1362,\\
Commission of the European Communities (DG XII), \\
Direction des Sciences de la Mati$\grave{\mbox{\rm e}}$re, CEA, France, \\
Bundesministerium f$\ddot{\mbox{\rm u}}$r Bildung, Wissenschaft, Forschung 
und Technologie, Germany,\\
General Secretariat for Research and Technology, Greece, \\
National Science Foundation (NWO) and Foundation for Research on Matter (FOM),
The Netherlands, \\
Norwegian Research Council,  \\
State Committee for Scientific Research, Poland, SPUB-M/CERN/PO3/DZ296/2000,
SPUB-M/CERN/PO3/DZ297/2000 and 2P03B 104 19 and 2P03B 69 23(2002-2004)\\
JNICT--Junta Nacional de Investiga\c{c}\~{a}o Cient\'{\i}fica 
e Tecnol$\acute{\mbox{\rm o}}$gica, Portugal, \\
Vedecka grantova agentura MS SR, Slovakia, Nr. 95/5195/134, \\
Ministry of Science and Technology of the Republic of Slovenia, \\
CICYT, Spain, AEN99-0950 and AEN99-0761,  \\
The Swedish Natural Science Research Council,      \\
Particle Physics and Astronomy Research Council, UK, \\
Department of Energy, USA, DE-FG02-01ER41155, \\
EEC RTN contract HPRN-CT-00292-2002. \\

\begin{figure}[tbhp]
\centerline{
\epsfxsize=16.0cm\epsffile{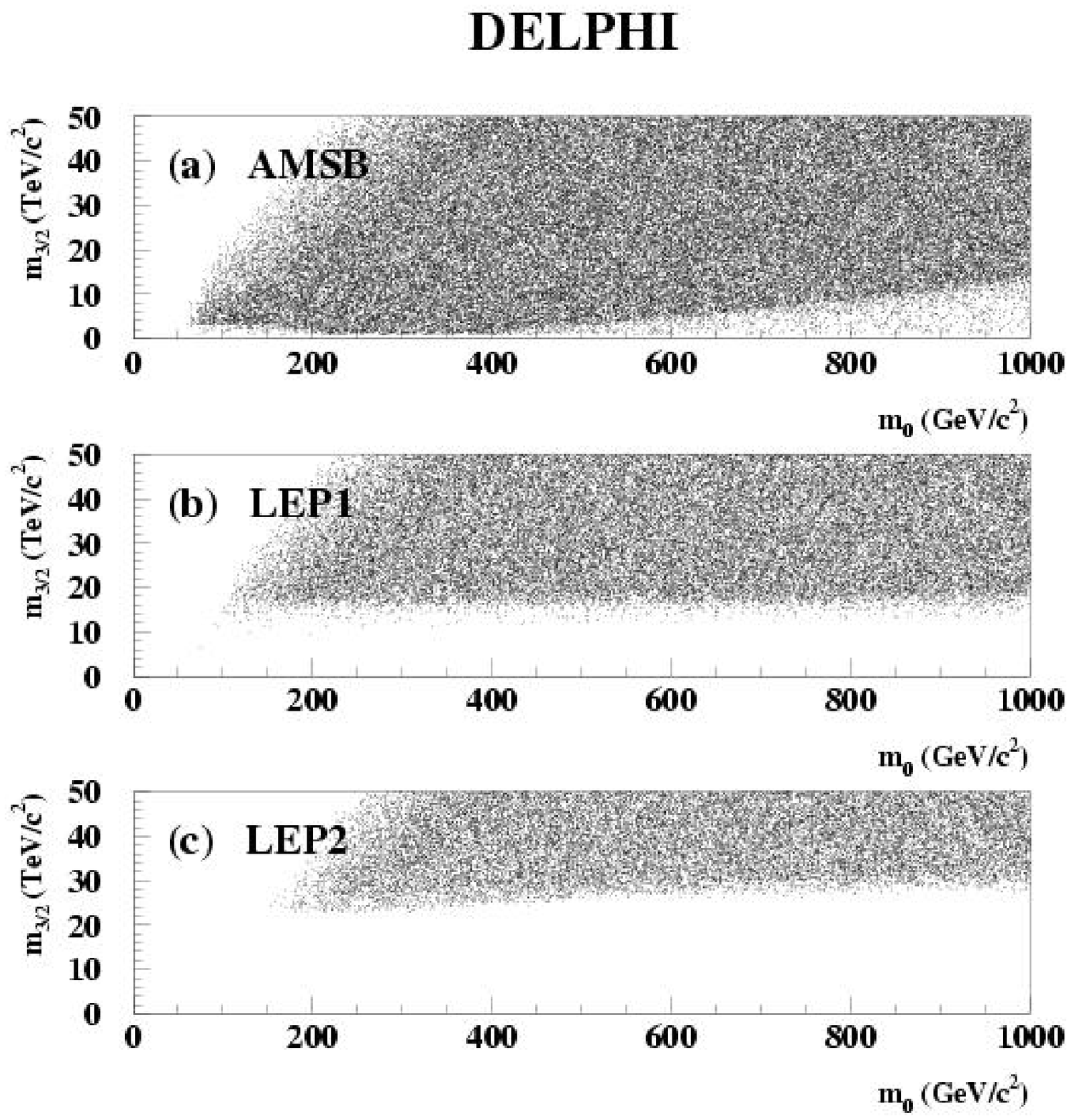} }
\caption[]{ (a) physically allowed \mzero\ and \mhf\ parameters in AMSB, as obtained in a scan
        of the AMSB parameter space with ISAJET, as described in the text. (b) points remaining
        after applying the chargino and sneutrino mass bounds of LEP1. (c) set of points from the
        scan remaining after considering all the results of the searches described in this work.
       }
\label{fig:isajet-mm}
\end{figure}

\begin{figure}[tbhp]
\centerline{
\epsfxsize=16.0cm\epsffile{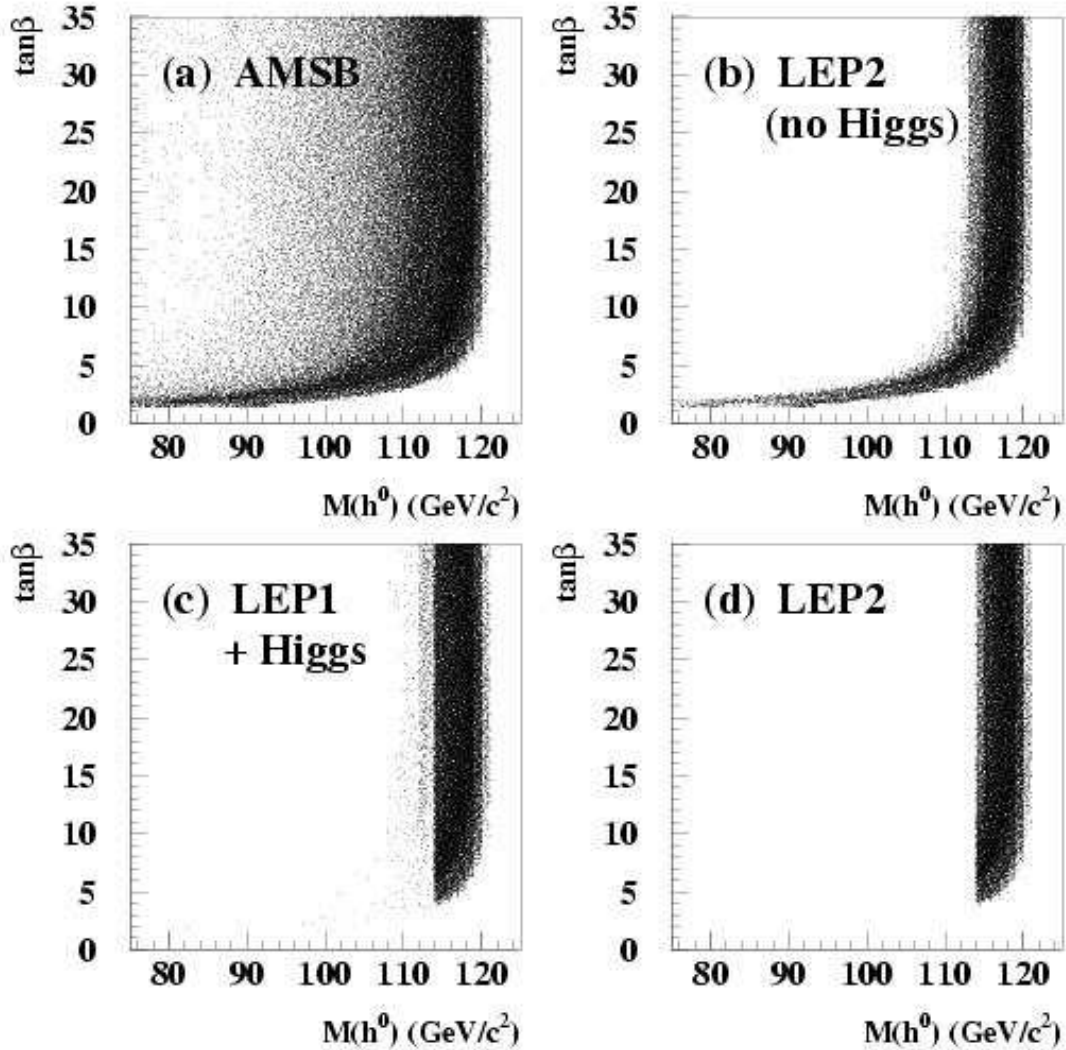} }
\caption[]{ (a) physically allowed \MSH\ and \tanb\ in AMSB, as obtained in a scan of the AMSB
        parameter space with ISAJET, as described in the text. (b) points remaining after applying
        the chargino and sneutrino mass bounds of LEP1 and the LEP2 search for SUSY particles, but
        the Higgses. (c) points remaining after applying the chargino and sneutrino mass bounds of
        LEP1 and the negative results of the searches for the SM and invisibly decaying Higgs
        bosons. (d) set of points remaining finally after considering all the results of the
        searches described in this work. No points survived for which $\MSH<114.1$~\GeVcc.
       }
\label{fig:isajet-hig}
\end{figure}

\end{document}